\documentclass[twocolumn,showpacs,preprintnumbers,amsmath,amssymb]{revtex4}
\usepackage{graphicx}% Include figure files
\usepackage{dcolumn}% Align table columns on decimal point
\usepackage{bm}% bold math
\usepackage{epsfig}
\usepackage{epstopdf}
\usepackage{grffile}
\usepackage{color}
\usepackage{colordvi}
\usepackage{amsmath,amssymb}
\usepackage{rotating}
\usepackage{lscape}
\usepackage{float}

\usepackage{footnote}
 \makesavenoteenv{environmentname}

\usepackage{hyperref}

\usepackage[T1]{fontenc} % if needed

\begin{document}

\title{QCD analysis of polarized DIS and the SIDIS asymmetry world data and light sea-quark decomposition }

\author{F. Arbabifar}

\email{farbabifar@ipm.ir}

\author{Ali N. Khorramian}

\email{khorramiana@theory.ipm.ac.ir}

\author{M. Soleymaninia}

\email{maryam_soleimannia@ipm.ir}

\affiliation{ Physics Department, Semnan University, Semnan, Iran \\
  School of Particles and Accelerators, Institute for
Research in Fundamental Sciences (IPM), P.O.Box 19395-5531, Tehran,
Iran  }

\date{\today}

\begin{abstract}The results of our new QCD analysis of helicity parton
distributions of the nucleon at full next-to-leading order (NLO)
accuracy in the fixed-flavor number scheme will be presented.
Performing a combined QCD fit on the global sets of latest inclusive
and the
 semi-inclusive polarized deep inelastic scattering data, we are
able to extract new polarized parton distribution functions (PPDFs)
at the input scale $Q_{0}^2=1$~GeV$^2$. Particulary, we have
calculated PPDFs considering light sea-quark decomposition
  and the results are compared with the experimental data and the most precise theoretical
 models obtained by recent analyses. The latest COMPASS2010 SIDIS data, which were not available for the previous analyses,
  are employed in the current analysis
 and the effect of COMPASS SIDIS data is studied in detail. Also the
uncertainties of PPDFs are determined using the standard Hessian
technique.

\end{abstract}

\maketitle

\section{Introduction}
In the recent years the determination of nucleon partonic
composition and their spin projections from high energy experimental
data has improved remarkably and the extracted polarized and
unpolarized partonic distributions have very essential role in the
study of hard scattering processes phenomenology. For the case of
polarized parton distributions, the experimental discoveries of
nucleon spin in the late 1980's \cite{EMC1,EMC2} proved the spin
contribution of valence quarks is anomalously small and the
predictions are far from reality, then the theoretical assumptions
on perturbative QCD were applied to interpret theses experimental
results
\cite{Anselmino:1994gn,Lampe:1998eu,Hughes:1999wr,filippone-01,Altarelli:2009za}.

 In the recent decades the
determination of PPDFs and their uncertainties from deep inelastic
scattering (DIS) experiments performing at CERN, SLAC, DESY and JLAB
\cite{EMCp,SMCpd,COMP1,E142n,E143pd,E154n,E155d,E155p,hermespdf,HERMn,HERMpd,JLabn,CLA1pd}
spread very fast
\cite{Ball:2013lla,Gehrmannn,Blumlein:2010rn,PRD11,lea99,Ball:1997sp,
bou98,gor98,alt98abfr,str99,gho00,
Gluck:2000dy,Bhalerao:2001rn,Leader:2001kh,
Bluemlein:2002be,Goto:1999by,Leader:2005ci,
Bourrely:2001du,ABFR,alt97, Hirai:2008aj,
Khorramian:2004ih,AtashbarTehrani:2007be,
AtashbarTehrani:2011zz,Arbabifar:2011zz,Khorramian:2010zz} and
recently semi inclusive deep inelastic scattering (SIDIS)
experimental data \cite{SMC98,hermespdf,COMP1_h,COMP1_pK} have been
also included by some of the theoretical groups \cite{lss, dns,
deFlorian:2000bm, deFlorian:2005mw, deFlorian:2008mr}. The only
theoretical group which perform a combined NLO analysis of DIS,
SIDIS and polarized proton proton collision data was DSSV in 2008
\cite{deFlorian:2009vb}. The extracted PPDFs of valence quarks
lightly differ but the PPDFs of sea quarks and gluon are more
different caused by datasets selection, parametrization forms of
PPDFs and the method of evolution and QCD analysis. The effect of
different PPDFs and the spin Physics on the determination of
fragmentation functions have been studied recently in
Ref.~\cite{Soleymaninia:2013cxa}

 In our latest
analysis we studied the impact of inclusive DIS data on the
determination of PPDFs based on Jacobi polynomials with flavor
symmetric light sea distribution, i.e.
$\delta\bar{u}=\delta\bar{d}=\delta\bar{s}=\delta s$ \cite{PRD11},
and now we consider light sea-quark decomposition and include
additional SIDIS data
\cite{SMC98,hermespdf,COMP1_h,COMP1_pK,compass10}. In fact, fully
inclusive DIS data from many different experiments are just
impressive to determine the sum of the quark and anti-quark
distributions while SIDIS data help to tell difference between
quarks and anti-quarks as well. Here we focus on the effect of SIDIS
data on determination of PPDFs, specially sea quarks distribution
separation which was not considered in our last DIS data analysis
and we present the comparison between both results. The impact of
RHIC polarized proton proton collision data will be studied in a
separate publication in near future.

 In the present analysis we utilize the full sets of proton and deuteron SIDIS asymmetry
data from the COMPASS group at CERN
\cite{COMP1_h,COMP1_pK,compass10} which were partially considered by
the last analysis on SIDIS data \cite{deFlorian:2009vb,lss}.
Specially we use the new semi inclusive asymmetries COMPASS2010
proton data for charged pions and kaons production from polarized
proton target,
 $A_{1}^{p,\pi^{\pm}}$ and $A_{1}^{p,k^{\pm}}$ \cite{compass10}, which were not
available for the analysis before 2010 and are helpful to study
$\delta s$ and $\delta \bar{s}$ distributions due to kaon detection
from polarized proton for the first time \cite{compass10}. In order
to discuss more about the effect of COMPASS SIDIS data on polarized
$\bar{u}$, $\bar{d}$ and $s=\bar{s}$ we perform extra analysis
excluding these datasets. The comparison of results shows the effect
of their inclusion clearly.

 The current study presents a new NLO QCD analysis of the polarized DIS and SIDIS
  data and we extract new
parametrization forms of PPDFs in flavor SU(2) and SU(3) symmetry
breaking. Here we also propose a simplified form of double Mellin
convolution expressions which saves times during fitting procedure.
Also the behavior of $\Delta\chi^2$ and the uncertainty of PPDFs are
calculated using the standard Hessian method.

This paper is organized as follows. In Sec.~\ref{dataset} we present
the relationship between polarized structure functions and asymmetry
data as observables and then we review the datasets used in our
analysis on PPDFs. QCD analysis including parametrization, evolution
and simplification of double Mellin convolution of PPDFs, Wilson
coefficients and FFs are discussed in Sec.~\ref{pdfsec}.
Sec.~\ref{fit} presents the fitting procedure and global $\chi^2$
minimization for asymmetry data and the investigation of $\chi^2$
neighborhood  and error calculation by standard Hessian method. We
present the full results of our fit to the data, comparison with
other models and sum rules in Sec.~\ref{results} and finally
Sec.~\ref{summary} contains the summary of the whole work.
\vspace{-0.75cm}
\section{Experimental observables and datasets}\label{dataset}
\subsection{Polarized asymmetries}
Perturbative QCD can predict polarized structure function
$g_1(x,Q^2)$ in terms of PPDFs and strong coupling constant up to
NLO approximation. However, experimental groups measure cross
section asymmetries $A_{\parallel}$ and $A_{\bot}$ defined by the
following ratios
\begin{eqnarray}
A_{\parallel}&=&\frac{d\sigma^{\rightarrow\Rightarrow}-d\sigma^{\rightarrow\Leftarrow}}{d\sigma^{\rightarrow\Rightarrow}+d\sigma^{\rightarrow\Leftarrow}}~,\nonumber\\
A_{\bot}&=&\frac{d\sigma^{\rightarrow\Uparrow}-d\sigma^{\rightarrow\Downarrow}}{d\sigma^{\rightarrow\Uparrow}+d\sigma^{\rightarrow\Downarrow}}~,
\label{Asym}
\end{eqnarray}
where $d\sigma^{\rightarrow\Rightarrow}$ and
$d\sigma^{\rightarrow\Leftarrow}$ are cross sections for
longitudinal polarized lepton scattering of a parallel or
anti-parallel polarized target hadron, and
$d\sigma^{\rightarrow\Uparrow}$ and
$d\sigma^{\rightarrow\Downarrow}$ are the same for a transversely
polarized hadron.

The ratio of polarized and unpolarized structure functions, $g_1$
and $F_1$, is related to the measurable asymmetries by
\begin{eqnarray}
\frac{g_{1}(x,Q^2)}{F_1(x,Q^2)}&=&\frac{1}{(1+\gamma^2)(1+\eta\zeta)}\nonumber\\
&&\left[(1+\gamma\zeta)
\frac{A_{\parallel}}{D}-(\eta-\gamma)\frac{A_{\bot}}{d}\right]~,
\label{g1F1ratio}
\end{eqnarray}
and the definitions of kinematic factors are given by
\begin{eqnarray}
\gamma&=&\frac{2Mx}{\sqrt{Q^2}}~,\\
d &=&\frac{D\sqrt{1-y-\gamma^2y^2/4}}{1-y/2}~,\\
D &=&\frac{1-(1-y)\epsilon}{1+\epsilon R(x,Q^2)}~,\\
\eta &=&\frac{\epsilon \gamma y}{1-\epsilon(1-y)}~,\\
\zeta &=&\frac{\gamma (1-y/2)}{1+\gamma^2 y/2}~,\\
\epsilon &=&\frac{4(1-y)-\gamma^2 y^2}{2y^2+4(1-y)+\gamma^2 y^2}~,
\label{kinematics}
\end{eqnarray}
here $M$ denotes the nucleon mass and $y=(E-E')/E$ describes the
normalized energy fraction transferred to the virtual photon, with
$E$ the energy of incoming lepton and $E'$ the energy of scattered
lepton in the nucleon rest frame. The unpolarized structure function
$F_1$ is expressed by its expression in terms of measured
unpolarized structure function $F_2$ extracted from unpolarized DIS
experiments \cite{NMC}
\begin{equation}
F_1(x,Q^2)= \frac{(1+\gamma^2)}{2x(1+R(x,Q^2))}F_2(x,Q^2)~,
\label{F2}
\end{equation}
and $R$ is the ratio of longitudinal to transverse photon-nucleon
unpolarized structure function which is determined in Ref.
\cite{R1998}
\begin{equation}
R(x,Q^2)= \frac{F_L(x,Q^2)}{F_2(x,Q^2)-F_L(x,Q^2)}~. \label{F2}
\end{equation}

The asymmetries $A_{\parallel}$ and $A_{\bot}$ can be expressed in
terms of $A_1$ and $A_2$, which are the virtual-photon longitudinal
and transverse asymmetries, by
\begin{eqnarray}
A_{\parallel}&=&D(A_1+\eta A_2)~,\nonumber\\
A_{\bot}&=&d(A_2-\zeta A_1)~, \label{Asym2}
\end{eqnarray}
where
\begin{eqnarray}
A_{1}(x,Q^2)&=&\frac{\sigma_{1/2}^{T}-\sigma_{3/2}^{T}}{\sigma_{1/2}^{T}+\sigma_{3/2}^{T}}~,\nonumber\\
A_{2}(x,Q^2)&=&\frac{2\sigma^{TL}}{\sigma_{1/2}^{T}+\sigma_{3/2}^{T}}~.
\label{Asym3}
\end{eqnarray}
Note that $\sigma_{1/2}^{T}$ and $\sigma_{3/2}^{T}$ recall the
virtual transversly polarized photon scattering cross sections when
the total spin of photon-nucleon system is $1/2$ or $3/2$
respectively, and $\sigma^{TL}$ is the term denoting the
interference of longitudinal and transverse photon-nucleon
amplitudes. Finally using Eqs. \ref{Asym2} and \ref{g1F1ratio} one
can find the relation between polarized and unpolarized structure
functions $g_1$ and $F_1$, and the asymetries $A_1$ and $A_2$
\cite{Bass:2004xa}
\begin{eqnarray}
\frac{g_1}{F_1}=\frac{1}{1+\gamma^2}[A_1+\gamma A_2]~. \label{Asym4}
\end{eqnarray}
The value of $A_2$ has been determined by SMC \cite{SMCR}, E154
\cite{E154R} and E143 \cite{E143R} and the measurements showed its
contribution to $g_1/F_1$ can be neglected in a good approximation,
also it is being suppressed by the small value of kinematic factor
$\gamma$ in the limit $m^2\ll Q^2$.

 In our QCD analysis we perform fit procedure on
$A_1$ or $g_1/F_1$ for DIS data
\begin{equation}
A_1(x,Q^2)= \frac{g_1(x,Q^2)}{F_1(x,Q^2)}(1+\gamma^2)~. \label{A1}
\end{equation}
Note that such a procedure is equivalent to a fit to $(g_1)_{exp}$,
but it is more precise than the fit to the $g_1$ data themselves
presented by the experimental groups because here the $g_1$ data are
extracted in the same way for all of the datasets.

 Unlike the
inclusive polarized deep inelastic scattering wherein $g_1$
structure function is measured by detecting only the final state
lepton, the particle detected in semi inclusive polarized deep
inelastic experiments are charged hadrons in addition to scattered
lepton. When the energy fraction of hadron, $z=E_h/E_{\gamma}$ is
large, the most possible occurrence of detected hadrons are
$\pi^{\pm}$ and $k^{\pm}$ which include struck quarks in their
valance state. The double-spin asymmetry in SIDIS experiments for
the production of hadron $h$ is
\begin{equation}
A_{1N}^h(x,z,Q^2)=\frac{g_{1N}^h(x,z,Q^2)}{F_{1N}^h(x,z,Q^2)}~.
\label{A1sidis}
\end{equation}

 The structure functions $g_1^h$ and $F_1^h$ are fully determined in terms of polarized
 and unpolarized distributions respectively up to NLO
 approximation and will be discussed in Sec~\ref{pdfsec}.
 Thus we will determine $g_1$ and $g_{1N}^{h}$ from Eqs.~\ref{A1}
 and \ref{A1sidis} in the analysis and extract polarized parton
 distribution functions.
 \subsection{The datasets and ranges}
 We utilize two types of datasets from DIS and SIDIS experiments which
 come from relevant experiments done at DESY, SLAC, JLAB and CERN. The datasets used in our QCD analysis
are summarized in Table~\ref{tabledata1}.
 These experiments have different targets, including protons,
 neutrons and deuterons, and also different detected hadrons, including
 $\pi^{\pm}$, $k^{\pm}$ and $h^{\pm}$, for SIDIS reactions.
 We also show the
number of data and the kinematic cuts on the experiments in
Table~\ref{tabledata1}, we exclude the data points which are in the
range of $Q^2<1$ from our analysis, since below $Q^2=1$ GeV$^2$
perturbative QCD is not reliable. The summary of observables
is as follows:\vspace{4mm}\\
 $\bullet$ \textbf{\small{EMCp, SMCpd, COMPASS, E142, E143, E154, E155
HERMES and JLAB DIS data}} \\ These experiments all determined $A_1$
except of E155 \cite{E155p,E155d} and HERMES \cite{HERMn} which
present $g_1/F_1$ and JLAB \cite{JLabn} present both measurements.
Since we consider different masses of nucleons in $\gamma$ in
Eq.~\ref{A1}, we distinguish these two data types.

\begin{figure}[ht]
%\vspace{1cm}
%\includegraphics[width=85mm]{datadis.eps}% Here is how to import EPS art
 \centerline{\psfig{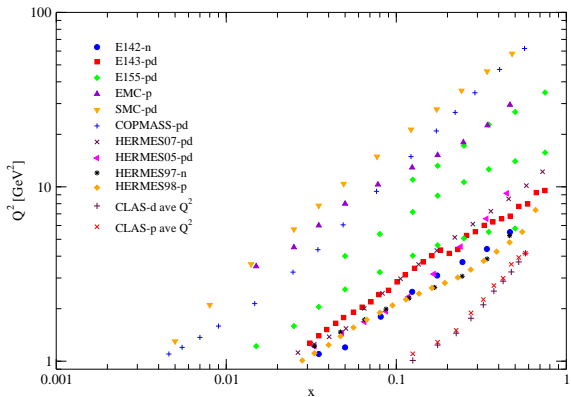}}
 \caption{DIS used data in a $(x,Q^2)$ plane.}\label{datadis}
\end{figure}

\begin{figure}[ht]
\vspace{1cm}
\centerline{\psfig{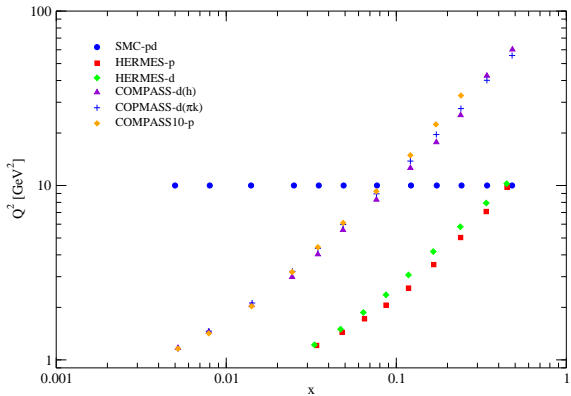}}
 \caption{SIDIS used data in a $(x,Q^2)$ plane.}\label{datasidis}
\end{figure}

 \noindent$\bullet$ \textbf{\small{SMCpd,
HERMES and COMPASS SIDIS
data}}\\
These experiments measure $A_{1N}^h$ as in Eq.~\ref{A1sidis} from
semi inclusive reactions. The very recent and precise proton data of
COMPASS \cite{compass10} are used for the first time in the current
analysis. Figs~\ref{datadis} and \ref{datasidis} shows the DIS and
SIDIS used data points in a scatter plot, as can been seen the
region of $x$ and $Q^2$ is restricted to $0.004\lesssim x \lesssim
0.75$ and $1.0\lesssim Q^2 \lesssim 60$ GeV$^2$ for DIS and
$0.005\lesssim x \lesssim 0.5$ and $1.0\lesssim Q^2 \lesssim 60$
GeV$^2$ for SIDIS experiments. We try to use all available data for
DIS and SIDIS experiments to cover a large range of kinematics
variables in comparison to other recent analysis
\cite{Blumlein:2010rn,lss,deFlorian:2009vb}.
\begin{table*}[!ht]
\hspace{-10mm}
 \centering \footnotesize{
 {\begin{tabular}{|c|c|c|c|c|c|c|c|c|c|c|c|c|c|c|c|c|}
\hline
 Experiment &Process &$N_{data}$ &$x_{min}$& $x_{max}$& $Q_{min}^{2}$~[GeV$^2$]&$Q_{max}^{2}$~[GeV$^2$]&$F$& $\chi^2$\\ \hline
 EMC \cite{EMCp}  &    DIS(p)   &  ~10 &0.015&0.466&3.5&29.5&$A_1^p$& 3.9 \\
 SMC \cite{SMCpd}  &    DIS(p)   &  ~12 &0.005&0.48&1.3&58&$A_1^p$ &3.4 \\
 SMC \cite{SMCpd}  &    DIS(d)   &  ~12 &0.005&0.479&1.3&54.8&$A_1^d$&  17.1  \\
 COMPASS \cite{COMP1} & DIS(p) &  ~15&0.0046&0.568&1.1&62.1&$A_1^p$& 20.5  \\
 COMPASS \cite{COMP1} & DIS(d) &  ~15&0.0046&0.566&1.1&55.3&$A_1^d$ &13.6  \\
 SLAC/E142 \cite{E142n}& DIS(n) &  ~~8&0.035&0.466&1.1&5.5&$A_1^n$ &4.18 \\
 SLAC/E143 \cite{E143pd}& DIS(p) &  ~28&0.031&0.749&1.27&9.52&$A_1^p$&22.0   \\
 SLAC/E143 \cite{E143pd}& DIS(d) &  ~28&0.031&0.749&1.27&9.52&$A_1^d$&54.6  \\
 SLAC/E154 \cite{E154n}& DIS(n) &  ~11&0.017&0.564&1.2&15&$A_1^n$  &3.3 \\
 SLAC/E155 \cite{E155p}& DIS(p) & ~24&0.015&0.75&1.22&34.72& $\frac{g_1^p}{F_1^p}$ &22.5 \\
 SLAC/E155 \cite{E155d}& DIS(d) & ~24&0.015&0.75&1.22&34.72& $\frac{g_1^d}{F_1^d}$&21.4   \\
 HERMES \cite{hermespdf} & DIS(p)  & ~~9&0.033&0.447&1.22&9.18&$A_1^p$&4.5  \\
 HERMES \cite{hermespdf} & DIS(d)  & ~~9&0.033&0.447&1.22&9.16&$A_1^d$ &11.4   \\
 HERMES \cite{HERMn} & DIS(n)  & ~~9&0.033&0.464&1.22&5.25&$A_1^n$&2.5   \\
 HERMES \cite{HERMn} & DIS(p)  & ~19&0.028&0.66&1.01&7.36&$\frac{g_1^p}{F_1^p}$ & 21.4 \\
 HERMES \cite{HERMpd} & DIS(p)  & ~15&0.0264&0.7248&1.12&12.21&$A_1^p$&10.2   \\
  HERMES \cite{HERMpd} & DIS(d)  & ~15&0.0264&0.7248&1.12&12.21&$A_1^d$  &16.7 \\
 JLab-Hall A \cite{JLabn}& DIS(n)& ~~3&0.33&0.6&2.71&4.38&$\frac{g_1^n}{F_1^n}$ &0.8 \\
 CLAS \cite{CLA1pd} &  DIS(p) & 151&0.1088&0.5916&1.01&4.96&$A_1^p$&151.0   \\
 CLAS \cite{CLA1pd} &  DIS(d) & 482&0.1366&0.57&1.01&4.16&$A_1^d$ &442.5  \\
\hline
 SMC \cite{SMC98} & SIDIS(p,$h^{+}$)& ~12&0.005&0.48&10&10&$A_1^{p,h^+}$&23.0  \\
 SMC \cite{SMC98} & SIDIS(p,$h^{-}$)& ~12&0.005&0.48&10&10&$A_1^{p,h^-}$& 11.9 \\
 SMC \cite{SMC98} & SIDIS(d,$h^{+}$)& ~12&0.005&0.48&10&10&$A_1^{d,h^+}$ &6.3 \\
 SMC \cite{SMC98} & SIDIS(d,$h^{-}$)& ~12&0.005&0.48&10&10&$A_1^{d,h^-}$ &17.2 \\
HERMES \cite{hermespdf}&SIDIS(p,$h^{+}$)& ~~9 &0.034&0.448&1.21&9.76&$A_1^{p,h^+}$&15.0 \\
HERMES \cite{hermespdf}&SIDIS(p,$h^{-}$)& ~~9&0.034&0.448&1.21&9.76&$A_1^{p,h^-}$&6.0  \\
HERMES \cite{hermespdf}&SIDIS(d,$h^{+}$)& ~~9&0.033&0.446&1.21&9.61&$A_1^{d,h^+}$ &10.3  \\
HERMES \cite{hermespdf}&SIDIS(d,$h^{-}$)& ~~9&0.033&0.446&1.21&9.61&$A_1^{d,h^-}$ &8.9  \\
HERMES \cite{hermespdf}&SIDIS(p,$\pi^{+}$)& ~~9&0.033&0.449&1.22&10.46&$A_1^{p,\pi^+}$ &9.7 \\
HERMES \cite{hermespdf}&SIDIS(p,$\pi^{-}$)& ~~9&0.033&0.449&1.22&10.46&$A_1^{p,\pi^-}$&7.7  \\
HERMES \cite{hermespdf}&SIDIS(d,$\pi^{+}$)& ~~9&0.033&0.446&1.22&10.24&$A_1^{d,\pi^+}$ &18.2\\
HERMES \cite{hermespdf}&SIDIS(d,$\pi^{-}$)& ~~9&0.033&0.446&1.22&10.24&$A_1^{d,\pi^+}$&25.0  \\
HERMES \cite{hermespdf}&SIDIS(d,$k^{+}$)& ~~9 &0.033&0.447&1.22&10.26&$A_1^{d,k^+}$&10.3  \\
HERMES \cite{hermespdf}&SIDIS(d,$k^{-}$)& ~~9 &0.033&0.447&1.22&10.26&$A_1^{d,k^-}$&6.1   \\
COMPASS \cite{COMP1_h}& SIDIS(d,$h^{+}$)& ~12&0.0052&0.482&1.17&60.2&$A_d^{h^+}$&18.1  \\
COMPASS \cite{COMP1_h}& SIDIS(d,$h^{-}$)& ~12 &0.0052&0.482&1.17&60.2&$A_d^{h^-}$&20.2 \\
COMPASS \cite{COMP1_pK}& SIDIS(d,$\pi^{+}$)& ~10&0.0052&0.24&1.16&32.8&$A_d^{\pi^+}$&13.8  \\
COMPASS \cite{COMP1_pK}& SIDIS(d,$\pi^{-}$)& ~10&0.0052&0.24&1.16&32.8&$A_d^{\pi^-}$&14.6  \\
COMPASS \cite{COMP1_pK}& SIDIS(d,$k^{+}$)& ~10&0.0052&0.24&1.16&32.8&$A_d^{k^+}$&24.0  \\
COMPASS \cite{COMP1_pK}& SIDIS(d,$k^{-}$)& ~10&0.0052&0.24&1.16&32.8&$A_d^{k^-}$&14.4   \\
COMPASS10 \cite{compass10}& SIDIS(p,$\pi^{+}$)& ~12&0.0052&0.48&1.16&55.6&$A_p^{\pi^+}$ &15.7  \\
COMPASS10 \cite{compass10}& SIDIS(p,$\pi^{-}$)& ~12&0.0052&0.48&1.16&55.6&$A_p^{\pi^-}$& 11.2  \\
COMPASS10 \cite{compass10}& SIDIS(p,$k^{+}$)& ~12&0.0052&0.48&1.16&55.6&$A_p^{k^+}$&14.3   \\
COMPASS10 \cite{compass10}& SIDIS(p,$k^{-}$)& ~12 &0.0052&0.48&1.16&55.6&$A_p^{k^-}$&6.4  \\
\hline
{\bf \ TOTAL}: &        &   1149 &&&&& &1171.5  \\
\hline
\end{tabular}}}
\caption{ Published data points from experimental groups, the
process which they are extracted from, the number of them (with a
cut of $Q^{2}\geq1.0$ GeV$^{2}$), their kinematic range, the
measured observables and the $\chi^2$ values for each set.
}\label{tabledata1}
\end{table*}
\section{Determination of polarized PDFs from observables}\label{pdfsec}
\subsection{Theoretical framework}
The idea behind our present analysis is to extract the universal
polarized PDFs entering factorized cross sections by optimizing the
agreement between the measured asymmetries from DIS and  SIDIS
experiments, relative to the accuracy of the data, and corresponding
theoretical calculations, through variation of the shapes of the
polarized PDFs. Considering perturbative QCD, the structure function
$g_1(x,Q^2)$ can be written in NLO approximation as
 a Mellin convolution of the PPDFs, including gluon, with
the corresponding Wilson coefficient functions $\delta C_{q,g}$
\cite{Lampe:1998eu} by
\begin{eqnarray}
\label{eqg1} g_1(x,Q^2)&=&\frac{1}{2} \sum_{q,\bar{q}}^{n_f} e_q^2
\left\{ \left[1+\frac{\alpha_{s}}{2\pi}\delta C_{q}\right]\right.
 \otimes \delta q(x,Q^2) \nonumber\\
 &&+  \left.\frac{\alpha_{s}}{2\pi}\:2\delta C_{g}\otimes\delta g(x,Q^2)\right\}\
 ,
 \label{g1dis}
\end{eqnarray}
 here $e_q$ denotes the charge of the quark flavor and
$\{\delta q,\delta\bar{q},\delta g\}$ are the polarized quark,
anti-quark, and gluon distributions, respectively.

  For the SIDIS asymmetry of Eq.~\ref{A1sidis} we have the
following forms for polarized and unpolarized structure functions in
NLO approximation
\begin{eqnarray}
 g_{1N}^h(x,z,Q^2)&=&\frac{1}{2}\sum _{q,\bar{q}}
^{n_f}e_{q}^2\left\{\hspace{-0.4cm}\phantom{\int\limits_a^b}\left[\delta
q \left(1+\otimes \frac{\alpha_s(Q^2)}{2\pi} \delta C_{qq}\otimes\right)D_q^h\right.\right. \nonumber\\
&&+ \delta q\otimes\frac{\alpha_s(Q^2)}{2\pi} \delta C_{gq}^{(1)}\otimes D_g^h \nonumber\\
&&+ \left.\left.\hspace{-0.4cm}\phantom{\int} \delta
g\otimes\frac{\alpha_s(Q^2)}{2\pi} \delta C_{qg}^{(1)}\otimes
D_q^h\right](x,z,Q^2)\phantom{\int\limits_a^b}\hspace{-0.4cm}\right\}~,\nonumber\\
\label{g1h}
\end{eqnarray}
and
\begin{eqnarray}
 F_{1N}^h(x,z,Q^2)&=&\frac{1}{2}\sum _{q,\bar{q}}
^{n_f}e_{q}^2\left\{\hspace{-0.4cm}\phantom{\int\limits_a^b}\left[
q \left(1+\otimes \frac{\alpha_s(Q^2)}{2\pi}  C_{qq}\otimes\right)D_q^h\right.\right. \nonumber\\
&&+  q\otimes \frac{\alpha_s(Q^2)}{2\pi} C_{gq}^{(1)}\otimes D_g^h \nonumber\\
&&+ \left.\left.\hspace{-0.4cm}\phantom{\int} g\otimes
\frac{\alpha_s(Q^2)}{2\pi} C_{qg}^{(1)}\otimes
D_q^h\right](x,z,Q^2)\phantom{\int\limits_a^b}\hspace{-0.4cm}\right\}~,\nonumber\\
 \label{F1h}
\end{eqnarray}
where $\delta q$ and $q$ denote polarized and unpolarized parton
distributions, $\delta C_{ij}^{(1)}(x,z)$ and $C_{ij}^{(1)}(x,z)$,
$i,j=q,g$, are
 Wilson coefficient functions presented in Ref. \cite{FSV} in the $\rm
\overline{MS}$ scheme. Also $D_{q,\bar{q}}^h$, $D_g^h$ denote the
corresponding fragmentation functions and $n_f$ presents the number
of active flavors which we take $n_f=3$ in the present analysis.

 As can be seen, the SIDIS asymmetries depend on the hadronic
variable $z=p_h.p_N/p_N.q$ in addition to $x$ and $Q^2$. Here $z$
denotes the momentum fractions taking by the resulting hadron from
the scattered parton and $p_h$, $p_N$ and $q$ are the usual hadron,
nocleon and photon four momentum, respectively. Since experimental
collaborations do not present the $z$ variable of presented SIDIS
data points, we integrate over $z>0.2$, which comes from the current
fragmentation functions region, for both $g_1(x,z,Q^2)$ and
$F_1(x,z,Q^2)$ to cancel the $z$ dependance of $A_{1N}^h$
\cite{Sisakian:2000jn}
\begin{equation}
A_{1N}^h(x,z,Q^2)=\frac{\int_{0.2}^1 dz
g_{1N}^h(x,z,Q^2)}{\int^{1}_{0.2} dz F_{1N}^h(x,z,Q^2)}.
\label{A1sidisint}
\end{equation}

One of the most important ingredient of SIDIS data analysis is the
choice of fragmentation functions \cite{Leader:2012nh}. Although
there are different available analysis of FFs
\cite{Hirai:2007cx,Albino:2008fy,Soleymaninia:2013cxa}, here we use
the latest DSS \cite{DSS} NLO FFs and for unpolarized PDFs we choose
MRST02 \cite{Martin:2002aw} parametrization like DSSV09 and LSS10
\cite{deFlorian:2009vb,lss} to make our analysis comparable with
them. Also, in addition to the precision of above FFs and PDFs and
comparability , we use them together since DSS FFs were extracted
from SIDIS data using MRST02 unpolarized PDFs.

 Considering isospin
symmetry, one can relate proton and neutron parton distributions
\begin{eqnarray}
\delta u^p&=&\delta d^n~,~\delta \bar{u}^p=\delta \bar{d}^n~, \nonumber\\
\delta d^p&=&\delta u^n~,~\delta \bar{d}^p=\delta \bar{u}^n~, \nonumber\\
\delta s^p&=&\delta s^n~,~\delta \bar{s}^p=\delta \bar{s}^n,
\end{eqnarray}
so the polarized structure function of neutron $g_1^n$ can be
obtained from all of Eqs.~\ref{g1dis}, \ref{g1h} and \ref{F1h} by
just replacing up quarks PPDFs and FFs by down ones. Also deuteron
structure functions are given in terms of proton and neutron ones
\begin{eqnarray}
g_{1d}^{(h)}=\frac{1}{2}(g_{1p}^{(h)}+g_{1n}^{(h)})(1-1.5~\omega_D)~,
\end{eqnarray}
where $\omega_D=0.05$ is the probability to find the deuteron in a
$D$ state.
 Now by having PPDFs and all other ingredients, we are
able to make polarized asymmetry function from DIS and SIDIS
precesses.

\subsection{PPDFs Parametrization\label{par}}
In our analysis we choose an initial scale for the evolution of
$Q_{0}^2 = 1$~GeV$^2$ and assume the PPDFs to have the following
functional form
\begin{eqnarray}
x\:\delta q={\cal
N}_{q}\eta_{q}x^{a_{q}}(1-x)^{b_{q}}(1+c_{q}x^{0.5}+d_{q} x)~,
\label{eq:parm}\end{eqnarray}
 with  $\delta q=\delta u
+\delta\bar{u},~\delta d+\delta \bar{d},~\delta
\bar{u},~\delta\bar{d},~ \delta\bar{s}$ and $\delta g$. The
Normalization constants ${\cal N}_{q}$
\begin{eqnarray}
\frac{1}{{\cal
N}_{q}}&=&\left(1+d_{q}\frac{a_{q}}{a_{q}+b_{q}+1}\right)\,
B\left(a_{q},b_{q}+1\right) \nonumber\\
&&+
c_{q}B\left(a_{q}+\frac{1}{2},b_{q}+1\right)~,\label{eq:norm}\end{eqnarray}
 are
chosen such that $\eta_{q}$ are the first moments of $\delta
q(x,Q_{0}^{2})$ and $B(a,b)$ is the Euler beta function.
 \noindent Since the present SIDIS data are
not yet sufficient to distinguish $s$ from $\bar{s}$, we assume
$\delta{s}(x,Q^2)=\delta{\bar{s}}(x,Q^2)$ throughout.

To control the behavior of PPDFs, we have to consider some extra
constraints; so we get $a_{u+\bar{u}}=a_{\bar{u}}$ and
$a_{d+\bar{d}}=a_{\bar{d}}=a_s$ to control the small $x$ behavior of
$\bar{u}$, $\bar{d}$ and $s=\bar{s}$. Also in the primary fitting
procedures we find out that the parameters
$b_{\bar{u}},~b_{\bar{d}},~b_{s=\bar{s}}$ and $b_{g}$ become very
close to each other, around $10$. We understand that they are not
strongly determined by the fit, so we fix them to $10$ which is
their preferred value to fulfill the positivity condition,
$|\delta{q_i}(x,Q_o^2)|\leq{q_i}(x,Q_0^2)$ \cite{Soffer:1994ww}, and
also it controls the behavior of polarized sea quarks at large $x$
region. In addition, we find that the parameter $c_q$ is very close
to zero for $\delta q=\delta u +\delta\bar{u},~\delta d+\delta
\bar{d},~\delta\bar{s}$ and $\delta g$ so we fix them at $0$.

 Generally PPDFs analyses use
two well-known sum rules relating the first moments of PPDFs to $F$
and $D$ quantities which are evaluated in neutron and hyperon
$\beta$--decays \cite{PDG} under the assumption of SU(2) and SU(3)
flavor symmetries
\begin{eqnarray}
a_{3} & = & \Delta\Sigma_u -\Delta \Sigma_d=F+D\ ,\label{a3}\\
a_{8} & = & \Delta \Sigma_u+ \Delta \Sigma_d-2\Delta
\Sigma_{s}=3F-D\ ,\label{a8}\end{eqnarray}
 \noindent where $a_{3}$ and $a_{8}$ denote non-singlet combinations of the first
moments of the polarized parton distributions corresponding to
non-singlet $q_3$ and $q_8$ distributions
\begin{eqnarray}
q_{3} & = & (\delta u+\delta\overline{u})-(\delta d+\delta\overline{d})\ ,\\
q_{8} & = & (\delta u+\delta\overline{u})+(\delta
d+\delta\overline{d})-2(\delta s+\delta\overline{s})\ .
\end{eqnarray}

 A new reanalysis of $F$ and $D$ parameters with updated
$\beta$-decay constants acquired \cite{PDG} $F=0.464\pm0.008$ and
$D=0.806\pm0.008$, so we make use of these evaluations in our
present analysis; however, since we do not focus on flavor symmetry
and we have $\delta\bar{u}\neq\delta\bar{d}\neq \delta s$, we can
use the combination of Eqs. \ref{a3} and \ref{a8} as following
\begin{eqnarray}
\Delta u+\Delta \bar{u}&=&0.9275+\Delta s+\Delta\bar{s}~,\nonumber\\
\Delta d+\Delta \bar{d}&=&-0.3415+\Delta
s+\Delta\bar{s}~,\label{finalfirstmom}
\end{eqnarray}
and we apply the above relations in the analysis, so we exclude the
parameters define the first moment of $(\delta u+\delta \bar{u})$
and $(\delta d+\delta \bar{d})$ (i.e. $\eta_{u+\bar{u}}$ and
$\eta_{d+\bar{d}}$) from the analysis and obtain them by
Eq.~\ref{finalfirstmom}. The effect of symmetry breaking on the
first moment of PPDFs have been discussed in detail in the
literatures \cite{su3,deFlorian:2009vb}.

\subsection{Evolution \& Computational method \label{method}}
For numerical calculations we need the scale evolution of the PPDFs
from input scale $Q_0^2$ to each of the scales related to the data
points. This evolution is done by a well-known set of
integro-differential equations~\cite{ap,ap1} that can be easily
solved analytically after a transformation from $x$ space to Mellin
$N$-moment space. The Mellin transform of a generic function $f$
depending on momentum fraction $x$ is defined as
\begin{equation}
\label{mellint} f(N) \equiv \int_0^1 x^{N-1} f(x)\, dx\ .
\end{equation}

The transformation (\ref{mellint}) has the pleasant applied property
that convolutions change to ordinary products, which veritably
simplifies calculations based on Mellin moments
\begin{eqnarray}
[f\otimes g](N)\equiv \int_{0}^1 d x^{n-1} \int_z^1 \frac{dy}{y} f
\left( \frac{x}{y}\right)g(y)
 =f(N)g(N)~,\nonumber\\
 \label{conv}
\end{eqnarray}
it can be performed analytically not only for the relevant splitting
functions governing the evolution of the PDFs but even for the
partonic cross sections for both DIS and semi-inclusive DIS. The
Mellin transform of the parton distributions $q$ is defined similar
to Eq.~\ref{mellint}:
\begin{eqnarray}
\delta q(N,Q_{0}^{2})& =& \int_{0}^{1}x^{N-1}\:\delta
q(x,Q_{0}^{2})\:
dx \nonumber\\
 &= & \eta_{q}{\cal
N}_{q}\left(1+d_{q}\:\frac{N-1+a_{q}}{N+a_{q}+b_{q}}\right)\nonumber\\
&&\times B\left(N-1+a_{q},b_{q}+1\right) \nonumber\\
  &&+~c_q B\left(N+a_q-\frac{1}{2},b_q+1\right)~,\label{pdfn}
 \end{eqnarray}
\noindent where $q=\{u+\bar{u},d+\bar{d},\bar{u},\bar{d},s,g\}$ and
$B$ denotes the Euler beta function.

 The inverse Mellin transform reads
\begin{equation}
\label{eq:invmellin} f(x) \equiv \frac{1}{2\pi i} \int_{{\cal{C}}_N}
x^{-N} f(N)\, dN\ ,
\end{equation}
note that ${\cal{C}}_N$ is an appropriate contour in the complex $N$
plane which has an imaginary part with the range from $-\infty$ to
$+\infty$ and that crosses the real axis to the right of the
rightest pole of $f(N)$ \cite{Bluemlein:2002be}.

 Currently, in the case of using
asymmetry data of SIDIS \cite{deFlorian:2009vb,lss} which depends on
two scaling variables $x$ and $z$ according to Eqs.~\ref{g1h} and
\ref{F1h} , Mellin transformation and the inverse of that requires
straight extensions of Eqs.~\ref{mellint} and \ref{eq:invmellin} to
double transformations as was presented in
Refs.~\cite{ref:aemp,sv,dns}.

 Now for simplification of the double
convolution in Eq.~\ref{g1h} we apply a method to change it to a
single routine convolution by transforming the coefficients $\delta
C_{ij}(x,z)$ from $x-z$ space to $N-z$ space
\begin{eqnarray}
\int_0^1 x^{N-1}\delta C_{ij}(x,z)dx=\delta
C_{ij}(N,z)~,\label{MellinC}
\end{eqnarray}
and then we compute
\begin{eqnarray}
\int_{0.2}^{1}\delta C_{ij}(N,z)\otimes
D_{i(j)}(z)=\delta\widetilde{ C}_{ij}(N)~.
\end{eqnarray}

Finally we could have the evolution of NLO correction terms of
$g_{1N}^h$ in $N$-moment space according to Eqs.~\ref{g1h} and
\ref{conv}
\begin{eqnarray}
g_{1N}^h(N)&=&\frac{1}{2}\sum_{q,\bar{q}}^{n_f}
e_q^2\left\{\int_{0.2}^1 dz \delta q D_q^h \right.
+\frac{\alpha_s(Q^2)}{2\pi}\delta q
(N)\delta\widetilde{C}_{qq}(N)\nonumber\\
&&+\frac{\alpha_s(Q^2)}{2\pi}\delta q
(N)\delta\widetilde{C}_{gq}(N)\nonumber\\
&&+\left.\frac{\alpha_s(Q^2)}{2\pi}\delta g
(N)\delta\widetilde{C}_{qg}(N) \right\} ~,\label{g1n}
\end{eqnarray}
the process is the same for Eq.~\ref{F1h}. We calculate the
transformation of Eq.~\ref{MellinC} for all $C_{ij}(x,z)$ and
$\delta C_{ij}(x,z)$ and provide them in Appendix.
\section{Determination of minimum $\chi^2$ and errors}\label{fit}
\subsection{Minimization of $\chi^2$ }
The process of global QCD analysis is based on the minimization of
effective $\chi^2$ which shows the quality of the fit carried out on
datasets by variation of the input set of parameters.  We use the
QCD-PEGASUS program \cite{Vogt:2004ns} for the evolution of
distributions in $N$-moment space and the MINUIT package
\cite{MINUIT} for the minimization of $\chi^2$ function
\begin{equation}
\chi^{2}=\sum_{i}\left(\frac{\:
A_{1,i}^{exp}-A_{1,i}^{theor}}{\:\Delta A_{1,i}^{exp}}\right)^{2},
\end{equation}
 here $A_{1,i}^{exp}$,
$\Delta A_{1,i}^{exp}$, and $A_{1,i}^{theor}$ are the experimental
measured value, the experimental uncertainty and theoretical value
for the $i^{\mathrm{th}}$ data point, respectively. For the
experimental error calculation the statistical and systematic errors
of each data point are added in quadrature.
\begin{table*}[ht]
\centering  \footnotesize{
\begin{tabular}{|c|c|c|c|c|c|}
\hline
  flavor &  $\eta$  &  $a$ &  $b$ & $c$& $d$ \\ \hline
 $u+\bar{u}$& 0.783 & 0.409$ \pm$ 0.0025 & 2.733$\pm$ 0.0368
 &$0.0^*$&
80.855$\pm$1.4115 \\
 $d+\bar{d}$& -0.485 & 0.123$\pm$ 0.0036 & 4.249$\pm$ 0.0280
 &$0.0^*$&
83.345$\pm$13.9609 \\
$\bar{u}$& 0.051$\pm$ 0.0022 & 0.409$\pm$ 0.0025 &10.0$^*$  & 10.016$\pm$ 13.5510& -32.424$\pm$ 15.8386 \\
$\bar{d}$& -0.081$\pm$ 0.0020 & 0.123$\pm$ 0.0036& 10.0$^*$ &
116.235$\pm$81.2783&902.567$\pm$ 615.0900\\
$\bar{s}$ & -0.072$\pm$ 0.0077 & 0.123$\pm$ 0.0036 &10.0$^*$
&$0.0^*$&
-16.045$\pm$4.7815 \\
$g$ & -0.156$\pm$0.0039 & 2.453$\pm$ 0.0334& 10.0$^*$ &$0.0^*$& -3.922$\pm$0.0659 \\
\hline
\end{tabular}}
\caption{Final parameter values and their statistical errors at the
input scale $Q_0^2=1$ GeV$^2$, those parameters marked with (*) are
fixed. }\label{table2}
\end{table*}
 Currently present available
SIDIS data are not enough precise to determine strong coupling
constant at input scale, so according to the precise scale dependent
equation of $a_s=\frac{\alpha_s}{4\pi}$ used in PEGASUS in NLO
\cite{Vogt:2004ns}
\begin{eqnarray}
\frac{1}{a_s(Q^2)}&=&\frac{1}{a_s(Q_0^2)}+\beta_0\ln\left(\frac{Q^2}{Q_0^2}\right)\nonumber\\
&&- b_1
\ln\left\{\frac{a_s(Q^2)[1+b_1a_s(Q_0^2)]}{a_s(Q_0^2)[1+b_1a_s(Q^2)]}\right\}~,\label{as}
\end{eqnarray}
 we fixed $\alpha_{s}(Q_{0}^{2})=0.580$
which is corresponding to $\alpha_{s}(M_{Z}^{2})=0.119$, obtained
from MRST02 analysis \cite{Martin:2002aw}. In Eq.~\ref{as} we have
\begin{eqnarray}
\beta_0&=&11-\frac{2}{3} n_f~,\nonumber\\
 \beta_1&=&102-\frac{38}{3} n_f~, \nonumber\\
 b_1&=&\frac{\beta_1}{\beta_0}~.
\end{eqnarray}

Finally we minimize the $\chi^{2}$ with the 17 unknown parameters.
We work at NLO in the fixed-flavor number scheme $n_{f}=3$ in the
QCD evolution with massless partonic flavors.
\subsection{The neighborhood of $\chi^2_0$ and error determination via Hessian method \label{deltachi}}
Here we just present the essential points for studying the
neighborhood of $\chi^2_0$ and the full procedure is provided in
Refs.~\cite{Martin:2002aw,Martin:2009iq,Pumplin:2001ct}.
 As mentioned in Sec.~\ref{fit} we find the appropriate parameter set
which minimize the global $\chi^2$ function, we call this PDF set
$S_0$ and the parameters value of $S_0$, i.e. {$p_1^0....p_n^0$},
will be presented in Sec.~\ref{results}.

By moving away the
parameters from their obtained values, $\chi^2$ increases by the
amount of $\Delta \chi^2$
\begin{equation}
\Delta \chi^{2}=\chi^2-\chi^2_0=\sum_{i,j=1}^d H_{ij}
(p_i-p_i^0)(p_j-p_j^0)\label{delta chi}\ ,
\end{equation}
 where the Hessian matrix $H_{ij}$ is defined by
\begin{equation}
H_{ij}=\frac{1}{2}\frac{\partial^2\chi^2}{\partial p_{i}\partial
p_{j}}\mid_{min}\ ,
\end{equation}
and we note $C\equiv H^{-1}$. Now it is convenient to work in term
of eigenvalues and orthogonal eigenvectors of covariance matrix
\begin{equation}
\sum_{j=1}^n C_{ij}\upsilon_{jk}=\lambda_k \upsilon_{ik}\ ,
\end{equation}
also the displacement of parameter $p_i$ from its minimum $p_i^0$
can be expressed in terms of rescaled eigenvectors
$e_{ik}=\sqrt{\lambda_k}v_{ik}$
\begin{equation}
p_i-p_i^0=\sum_{k=1}^n e_{ik}z_k\ ,\label{zk}
\end{equation}
putting Eq.~\ref{zk} in \ref{delta chi} and considering the
orthogonality of $\upsilon_{ik}$ we have
\begin{equation}
\Delta\chi^2=\sum_{k=1}^{n}z_k^2\ .
\end{equation}

Now the relevant neighborhood of $\chi^2$ is the interior of
hypersphere with radius $T$:
\begin{equation}
\sum_{k=1}^nz_k^2\leq T^2\ ,
\end{equation}
and the neighborhood parameters are given by
\begin{equation}
p_i(s_k^\pm)=p_i^0\pm t e_{ik}\ ,\label{t}
\end{equation}
with $s_k$ is the $k^{th}$ set of PDF and $t$ adapted to make the
desired $T=(\Delta\chi^{2})^{\frac{1}{2}}$ and $t=T$ in the
quadratic approximation. In Sec.~\ref{results} we present the
dependance of $\Delta\chi^2$ along some random samples of
eigenvector directions to test the quadratic approximation of
Eq.~\ref{delta chi}.

 Now we accompany the construction of the QCD fit by reliable
estimation of uncertainty. As discussed in
Refs.~\cite{Martin:2002aw,Martin:2009iq,Pumplin:2001ct}, the master
equation to obtain the uncertainties of observables in modified
Hessian method is
\begin{equation}
\Delta
F=\frac{1}{2}\left[\sum_{k=1}^n(F(s_k^+)-F(s_k^-))^2\right]^{\frac{1}{2}}~,
\end{equation}
here $F(s_k^+)$ and $F(s_k^-)$ are the value of $F$ extracted from
the input set of parameters $p_i(s_k^\pm)$ instead of $p_i^0$
mentioned in Eq.~\ref{t}. However, it has been pointed out by DSSV
\cite{deFlorian:2009vb} that the modified Hessian method is known to
work reasonably well in extractions of spin independent parton
densities and it is found to fail in the case of helicity parton
densities for tolerances larger than $\Delta\chi^2=1$. So we prefer
to calculate the PPDFs error band using standard Hessian method wich
is more reliable in this case. As presented in Eqs.~\ref{pdfn} and
\ref{g1n}, the evolved polarized parton densities and structure
functions are attributive functions of the input parameters obtained
in the QCD fit procedure at the scale $Q_0^2$, then their
uncertainty can be written applying the standard Hessian method
\begin{equation}
\Delta F=\left[\Delta \chi^2 \sum_{i,j=1}^k \frac{\partial
F}{\partial p_i}C_{ij}\frac{\partial F}{\partial
p_j}\right]^{\frac{1}{2}}~.
\end{equation}
Here we calculate the PPDFs uncertainty with $\Delta\chi^2=1$ which
is the most appropriate choice in the polarized case. If one wishes
to choose $\Delta\chi^2>1$, one simply can scale our error bands by
$(\Delta\chi^2)^{1/2}$.

\section{Results\label{results}}
\subsection{The quality of QCD fit\label{fitquality}}
\begin{figure*}[ht]
\vspace{1cm} \centerline{\psfig{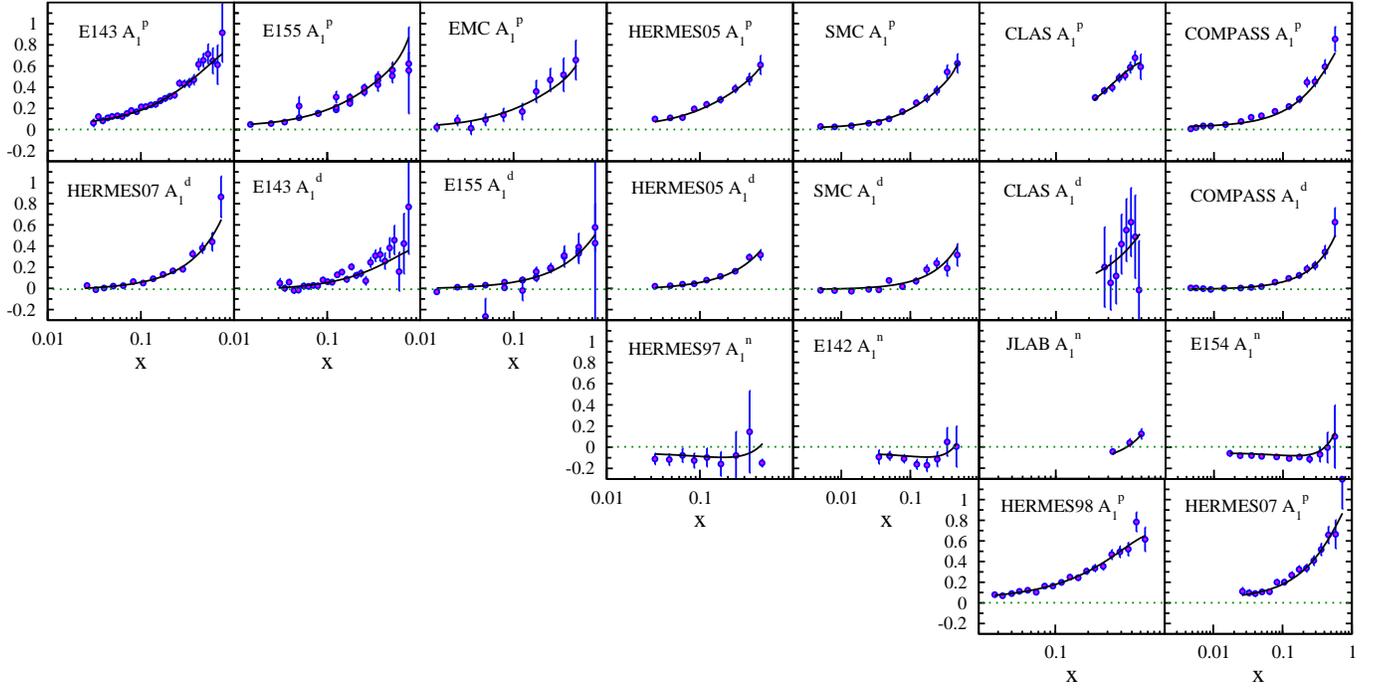}}
\caption{ Comparison of our NLO QCD results for the DIS asymmetries
of proton, neutron and deuteron with the data at measured $x$ and
$Q^2$.
 \label{QCDfitDIS1}}
\end{figure*}

\begin{figure*}[ht]
\vspace{1cm} \centerline{\psfig{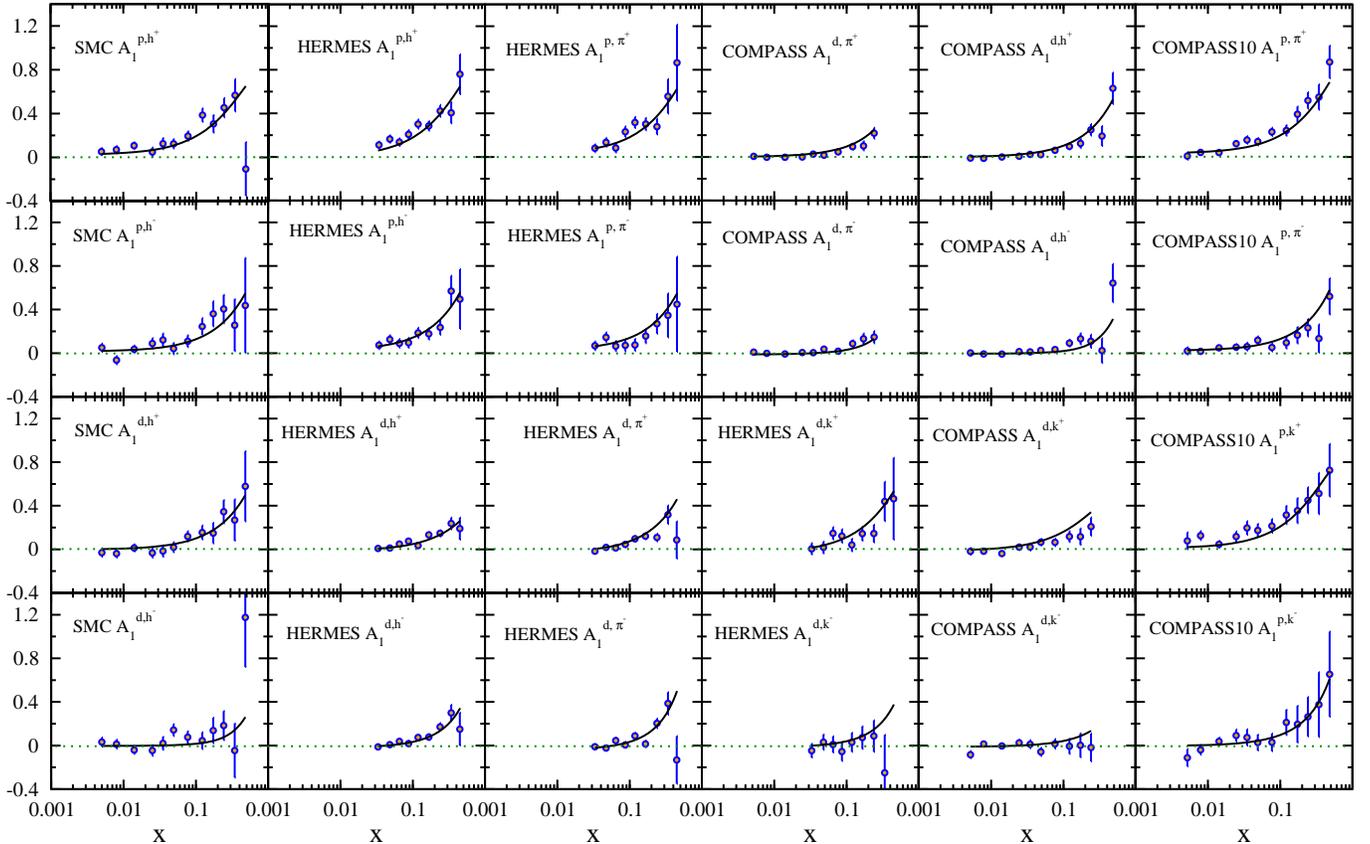}}
\caption{ Comparison of our NLO QCD results for the SIDIS
asymmetries with the data at measured $x$ and $Q^2$.
\label{QCDfitsidis2}}
\end{figure*}

The values of obtained parameters attached to the input PPDFs are
summarized in Table~\ref{table2}. We find ${\chi}^{2}/{\rm
{d.o.f.}}=1171.571/1132 = 1.03$ which yields an acceptable fit to
the experimental data, the individual $\chi^2$ for each set of data
is presented in Table.~\ref{tabledata1}. The quality of the QCD fit
to DIS and SIDIS asymmetry data is demonstrated in
Fig.~\ref{QCDfitDIS1} and Fig.~\ref{QCDfitsidis2}. Since we use many
number of CLAS DIS data for proton and deuteron \cite{CLA1pd} and
they do not situate in the Figure, we just show 8 of them for
presentation. As can be seen the data are generally well described
by the curves.
\subsection{Extracted polarized parton distributions}
\begin{figure*}[ht]
\vspace*{1cm}
\centerline{\psfig{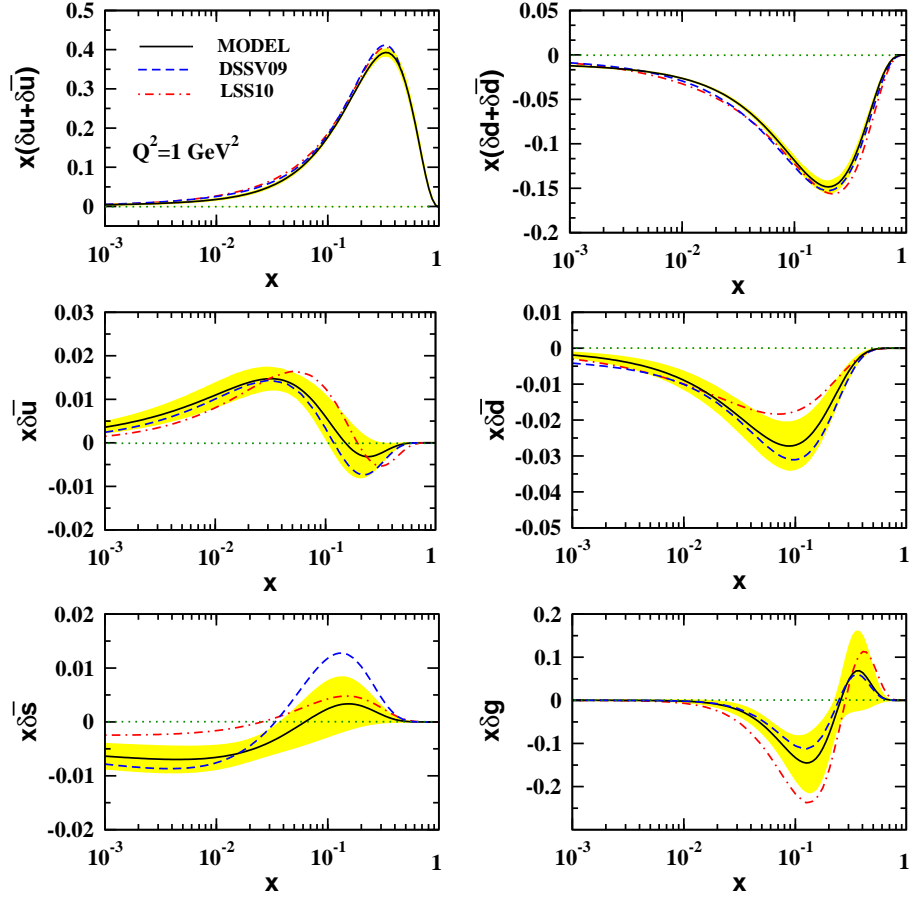}}
 \caption{The result of our analysis for quark helicity
distributions at $Q_{0}^2 = 1$ GeV$^2$ in comparison with DSSV09
\cite{deFlorian:2009vb} and LSS10 \cite{lss}.}\label{pdf}
\end{figure*}

In Fig.~\ref{pdf} we present the polarized parton distributions and
their comparison to parameterizations from DSSV09
\cite{deFlorian:2009vb} and LSS10 \cite{lss} at input scale
$Q_0^2=1$ GeV$^2$.

 Examining the $x(\delta u+\delta\bar{u})$ and
$x(\delta d+\delta\bar{d})$ distributions we see that all of the
fits are in agreement. For the $x\delta \bar{u}$ and $x\delta
\bar{d}$ distributions, the curves, specially our model and DSSV09,
are very close; $\delta\bar{d}$ is negative for any $x$ in the
measured $x$ region while $\delta\bar{u}$ passes zero around $x =
0.1-0.2$ and becomes negative for large $x$ for all presented
models.


\begin{figure}[h]
\vspace{1cm}
\centerline{\psfig{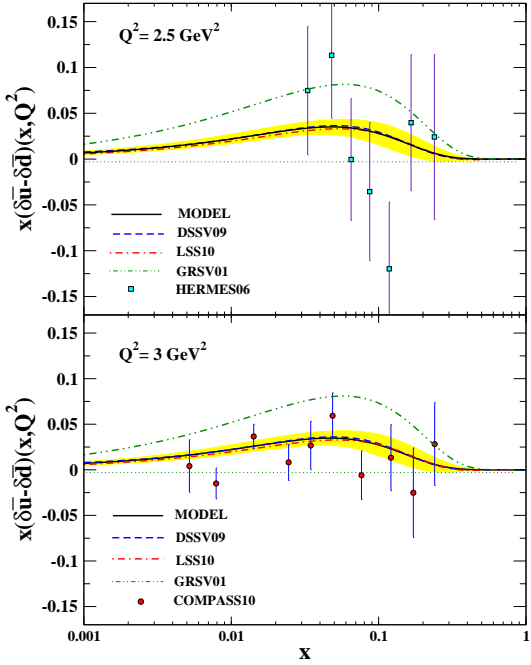}}
 \caption{The quark helicity
distributions for the difference $x(\delta\bar{u}-\delta\bar{d})$ at
$Q^2=2.5~,3$ GeV$^2$ comparing to DSSV09 \cite{deFlorian:2009vb},
LSS10 \cite{lss} and GRSV01 \cite{Gluck:2000dy} and experimental
data \cite{hermespdf,compass10}.}\label{mines}
\end{figure}

\begin{figure}[h]
\vspace{1cm}
\centerline{\psfig{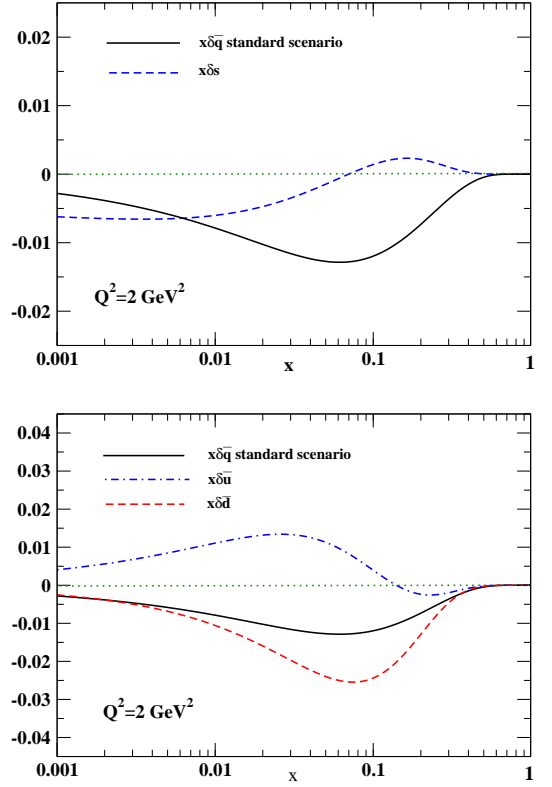}}
 \caption{The quark helicity
distributions for $x\delta s$, $x\delta\bar{u}$ and $x\delta\bar{d}$
at $Q_{0}^2 = 2$ GeV$^2$ comparing to $x\delta q$ obtained from the
previous standard scenario \cite{PRD11}.}\label{q}
\end{figure}

\begin{figure*}[ht]
\vspace{1cm}
\centerline{\psfig{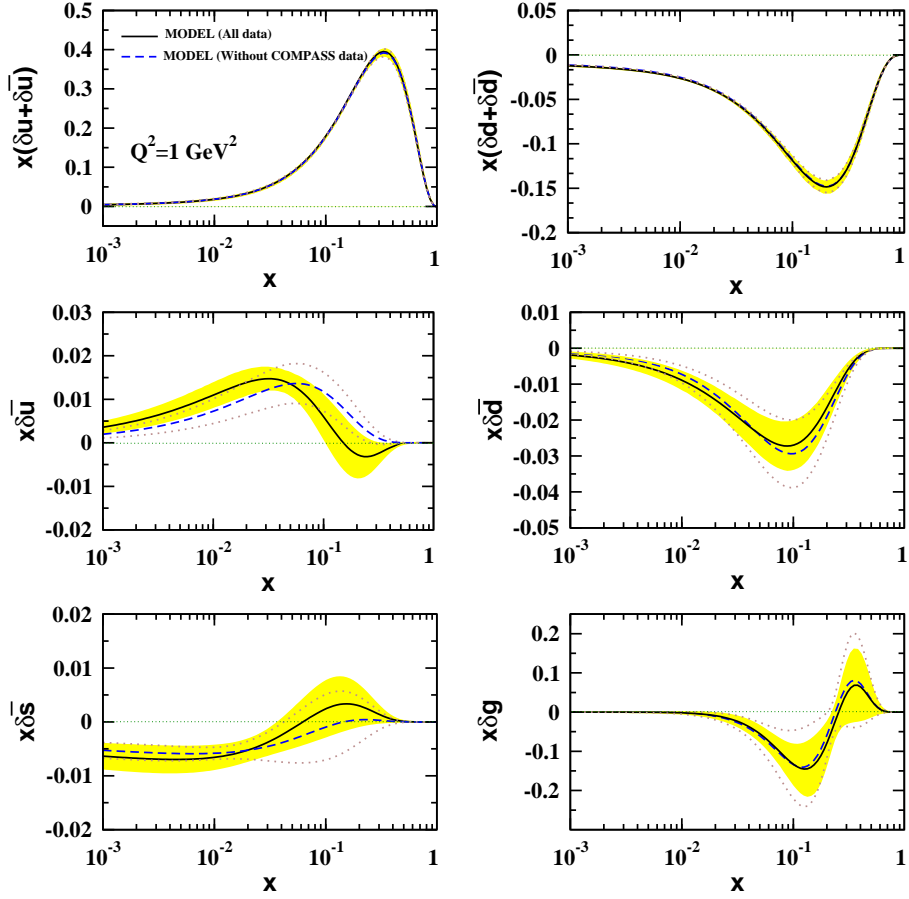}}
 \caption{The comparison of PPDFs results extracted from all
 datasets shown in Table~\ref{tabledata1}, and the PPDFs results extracted by excluding COMPASS SIDIS datasets
 at $Q_{0}^2 = 1$ GeV$^2$. The corresponding error bands of them are also shown.}\label{withoutcompass}
\end{figure*}

\begin{figure*}[ht]
\vspace{1cm}
\centerline{\psfig{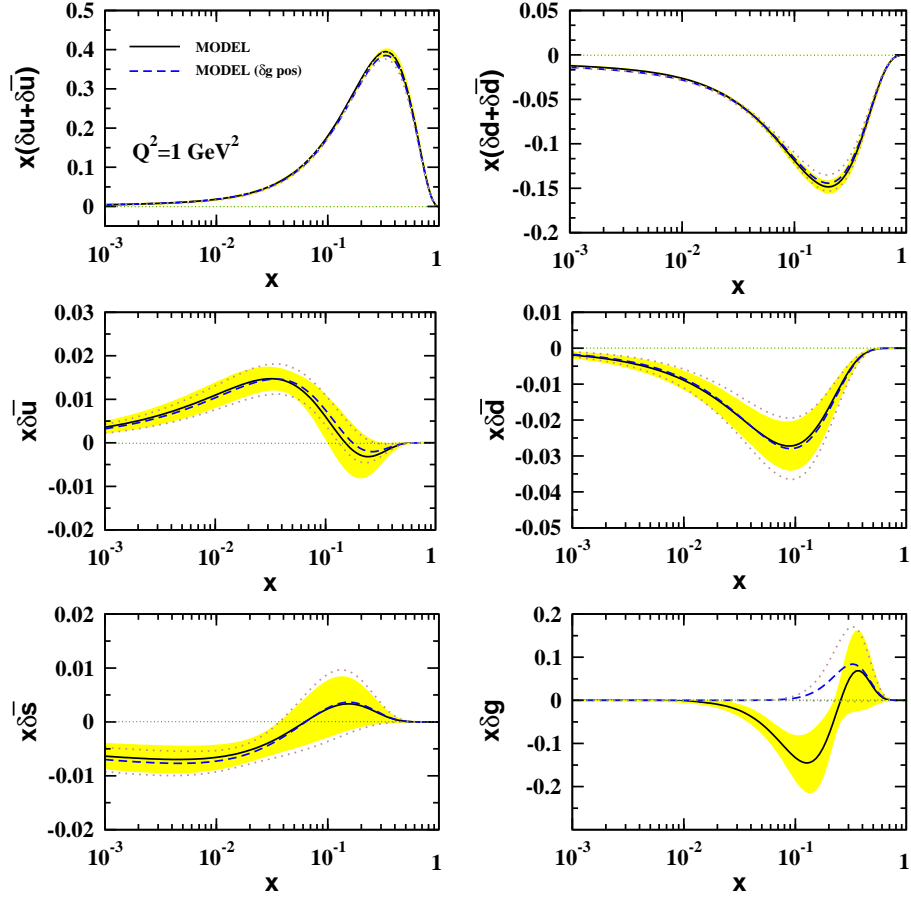}}
 \caption{The comparison of PPDFs results extracted from sign changing and positive gluon
scenarios at $Q_{0}^2 = 1$ GeV$^2$. The corresponding error bands of
them are also shown.}\label{posglu}
\end{figure*}

 For the strange sea-quark density $x\delta s$, the main difference between
the presented model, LSS10 and DSSV09 sets is that for $x < 0.03$
LSS10 is less negative than others, also both of current model and
LSS10 are less positive than DSSV09 for $x
> 0.03$. The other differences for the
distributions comes from the fact that both DSSV09 and LSS10
analysis uses different number of data (we use the most and the
newest ones) and DSSV09 uses pp collision data from RHIC which can
impose individual constraints and effects on individual parton
distributions in the nucleon \cite{deFlorian:2009vb}. As we
mentioned, in the current analysis we focuss on the study of SIDIS
data effect on determination of PPDFs, specially gluon and sea
quarks separation which was not considered in our last DIS analysis.
The impact of RHIC pp collision data will be considered in our near
future analysis.

\subsection{The impact of SIDIS data in determining the
polarized sea-quark distributions} Generally speaking, polarized
inclusive DIS data cannot distinguish $\bar{u}$, $\bar{d}$ and
$\bar{s}$ but $\delta u=\delta d=\delta s=
\delta\bar{s}=\delta\bar{q}$ is well determined and all the standard
scenario NLO QCD analysis yield a negative value of it for any $x$
in the measured region \cite{Blumlein:2010rn}. Employing SIDIS data
a flavor decomposition of the polarized sea quarks is obtained and
the light antiquark polarized densities $\delta\bar{u}$,
$\delta\bar{d}$ and $\delta s=\bar{s}$ are determined separately,
Fig.~\ref{mines} shows the difference between $\delta\bar{u}$,
$\delta\bar{d}$ in the current analysis comparing to other models
and experimental data.

Also in the present parametrization we use a term $(1 + c_s
x^{0.5}+d_s x)$ in the input strange sea-quark distribution to let a
sign changing for $\delta s=\delta\bar{s}$, which was not considered
in the standard scenario \cite{PRD11}. The comparison of polarized
light sea-quark distributions ($x\delta s,~x\delta \bar{u},~x\delta
\bar{d}$) in the standard scenario and current model are presented
in Fig.~\ref{q}. It shows that the behavior of the polarized strange
quark density remains puzzling \cite{Leader:2011tm}.

Note that by having SIDIS data $\delta s$ and $\delta \bar{s}$ can
be separately determined as was done by the COMPASS collaboration
recently \cite{compass10}. However, it was demonstrated that there
is no considerable difference between $\delta s$ and $\delta
\bar{s}$ in the $x$-range covered by their data. Also, the errors of
the presented values of the difference $\delta s(x)-
\delta\bar{s}(x)$ are quite large to allow us to conclude the
assumption $\delta s(x) = \delta\bar{s}(x)$ like LSS10 and DSSV09.
So, the above assumption and also the form of the fragmentation
functions used to extract PPDFs by different groups may be possible
causes of the contradiction between sea quarks densities obtained
from the analyses of inclusive DIS data and combined inclusive and
semi- inclusive DIS datasets \cite{Leader:2011tm}.
\subsection{The effect of COMPASS SIDIS data on
polarized sea-quark distributions} As shown in
Table~\ref{tabledata1} The measurement of SIDIS asymmetries for
unidentified charged hadrons was performed by SMC collaboration and
then the SIDIS asymmetries data for charged pion production from
proton target and for charged kaon and pion production from a
deuteron target was reported by HERMES. These asymmetries were used
in the most DIS and SIDIS QCD analysis but the SIDIS asymmetry data
from COMPASS are partially employed, specially the new semi
inclusive asymmetries COMPASS data for scattering of muons from a
polarized proton target for identified charged pions and kaons
production; $A_{1}^{p,\pi^{\pm}}$ and $A_{1}^{p,k^{\pm}}$
\cite{compass10}, which was not available for the analysis before
2010.

 In order to
study the effect of COMPASS SIDIS data on polarized parton
distributions, we show the comparison of our PPDFs results extracted
from all
 datasets shown in Table~\ref{tabledata1}, and the PPDFs results extracted by excluding COMPASS SIDIS datasets
 in Fig.~\ref{withoutcompass}. As can be seen COMPASS data
 has effect on $\delta\bar{u}$, $\delta\bar{d}$ and $\delta s=\delta\bar{s}$
distributions since $\pi^{\pm}$ and $k^{\pm}$ are directly related
to them. The changes of sea quark distributions are considerable in
$0.07\leq x\leq 0.4$ region that is well covered by the new COMPASS
data. The distributions of $\delta u +\delta\bar{u}$, $\delta
d+\delta\bar{d}$ and $\delta g$ are not changed considerably.
\subsection{Gluon polarization}In order to study the effect of SIDIS data on polarized gluon distribution, we
perform another analysis with positive polarized gluon distribution.
We understand that utilizing polarized DIS and SIDIS data in the QCD
analysis can not enforce the positive or sign changing polarized
gluon distributions and the data can not distinguish between these
two scenarios, although the recent analyses which employ both DIS
and SIDIS data (and not pp collision data) could extract sign
changing $x\delta g$ \cite{lss,Sissakian:2009ag}. In
Table.~\ref{posglutable} we present the values of obtained
parameters in positive gluon scenario. We find ${\chi}^{2}/{\rm
{d.o.f.}}=1.02$ which is almost equal to the one obtained from sign
changing gluon scenario ${\chi}^{2}/{\rm {d.o.f.}}=1.03$.
Fig.~\ref{posglu} shows the comparison of PPDFs extracted from
positive and sign changing gluon scenarios. As can be seen all the
distributions are almost not changed except $x\delta g$.

\begin{table*}[ht]
\centering  \footnotesize{
\begin{tabular}{|c|c|c|c|c|c|}
\hline
  flavor &  $\eta$  &  $a$ &  $b$ & $c$& $d$ \\ \hline
 $u+\bar{u}$& 0.770 & 0.411$ \pm$ 0.0955 & 2.760$\pm$ 0.1356
 &$0.0^*$&
78.145$\pm$39.9149 \\
 $d+\bar{d}$& -0.498 & 0.125$\pm$ 0.0080 & 4.221$\pm$ 0.0359
 &$0.0^*$&
67.151$\pm$4.0270 \\
$\bar{u}$& 0.051$\pm$ 0.0004& 0.411$\pm$ 0.0955 &10.0$^*$  & 11.231$\pm$ 0.4776& -32.425$\pm$ 0.6351 \\
$\bar{d}$& -0.081$\pm$ 0.0003 & 0.125$\pm$ 0.0080& 10.0$^*$ &
85.423$\pm$2.4403&915.221$\pm$ 26.7240\\
$\bar{s}$ & -0.079$\pm$ 0.0387& 0.125$\pm$ 0.0080 &10.0$^*$
&$0.0^*$&
-15.843$\pm$3.2887 \\
$g$ & 0.081$\pm$0.0008 & 3.958$\pm$ 0.0326& 10.0$^*$ &$0.0^*$& 25.820$\pm$0.8029 \\
\hline
\end{tabular}}
\caption{Final parameter values and their statistical errors at the
input scale $Q_0^2=1$ GeV$^2$ for positive gluon scenario, those
parameters marked with (*) are fixed. }\label{posglutable}
\end{table*}

\begin{figure}[pt]
\vspace{2cm}
\centerline{\psfig{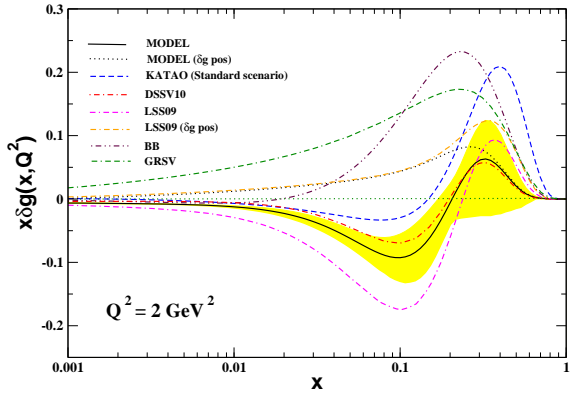}}
 \caption{The gluon helicity
distributions at $Q^2 = 2$ GeV$^2$ comparing to symmetry breaking
\cite{deFlorian:2009vb,lss} and standard scenario
\cite{PRD11,Gluck:2000dy,Blumlein:2010rn} analysis.}\label{glu}
\end{figure}

 Fig.~\ref{glu} shows the gluon distribution
comparison between our previous standard scenario and the present
sea flavor decomposition analysis. As presented, SU(2) and SU(3)
symmetry breaking has direct effect on $x\delta g(x,Q^2)$ and makes
it less. The gluon results of other models of both standard and
light sea-quark decomposition scenario are also presented in
Fig.~\ref{glu} and they confirm this effect too
\cite{deFlorian:2009vb,lss,PRD11,Gluck:2000dy,Blumlein:2010rn}.

\begin{figure}[ht]
%\hspace{-1cm}
 %\vspace{-1cm}
%\includegraphics[width=85mm]{deltag}
\centerline{\psfig{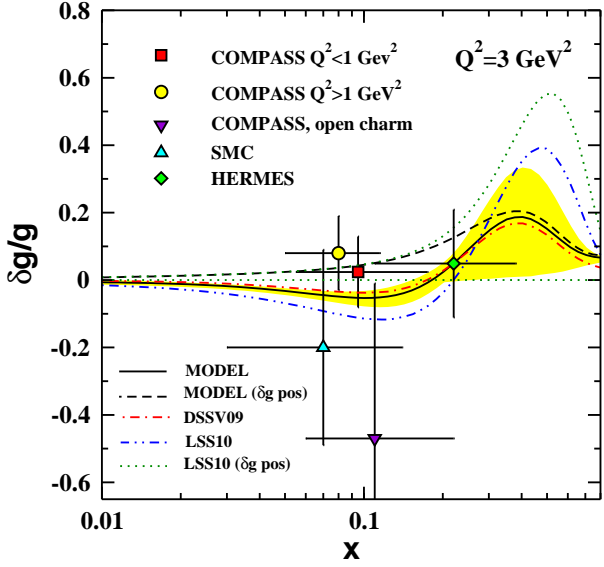}}
 \caption{The calculated ratio $\delta g/g$ for DSSV, LSS \cite{deFlorian:2009vb,lss} and our model in comparison with experimental data from COMPASS, HERMES and SMC
 \cite{Airapetian:1999ib,Adeva:2004dh,Stolarski:2008jc,Ageev:2005pq,Alekseev:2008cz} at  $Q^2=3$ GeV$^2$.
}\label{deltag}
\end{figure}

 We also calculated the ratio $\delta g/g$, using our extracted PPDFs
for polarized gluon distributions and MRST02 \cite{Martin:2002aw}
NLO QCD analysis for unpolarized gluon distributions. In
Fig.~\ref{deltag} we present a comparison for $\delta g/g$ from
LSS10, DSSV09 and our models (sing changing and positive gluon
scenarios) at $Q^2=3$ GeV$^2$ with measured value from experimental
data
\cite{Airapetian:1999ib,Adeva:2004dh,Stolarski:2008jc,Ageev:2005pq,Alekseev:2008cz}.
Although we do not use the mentioned data in the current analysis,
our results can predict the data very well.

\subsection{Behavior of $\Delta \chi^2$}
For more deliberation, in the preliminary QCD fit process we let all
input parameters to vary. While investigating the behavior of
$\Delta\chi^2$ we observe $\chi^2$ increase consumedly in some
points and
 a big amount of redundance in parameters happens, this redundancy
 results into disorder of quadratic behavior of $\Delta\chi^2$. In
 order to have the Hessian method work as shown in Sec.~\ref{deltachi}, we
 fix some parameters at their best obtained value so that Hessian
 matrix depends on the number of parameters which are independent sufficiently for
 the quadratic behavior of $\Delta\chi^2$, the detailed fixing and constraints in
 parameter space was discussed in Subsec.~\ref{par} and Sec.~\ref{fit}.

\begin{figure}[pt]
 %\hspace{-0.5cm}
\vspace{1.cm}
\centerline{\psfig{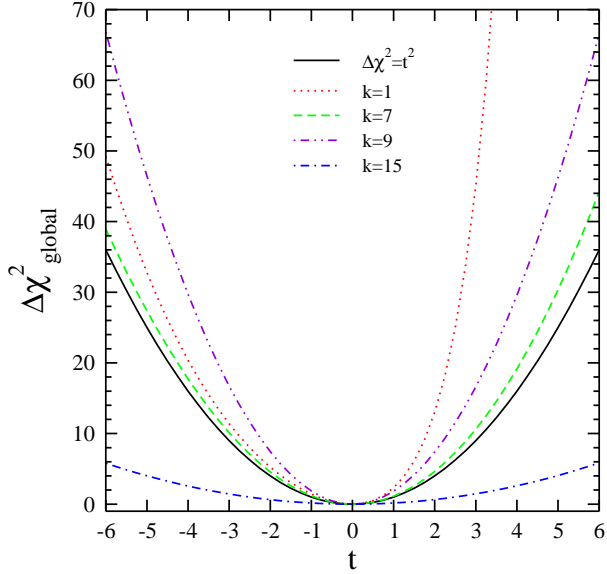}}
 \caption{$\Delta \chi^2_{global}$ as a function of $t$ for some random sample eigenvectors.}\label{deltachi2}
\end{figure}

 To test of quadratic approximation in Eq.~\ref{delta chi},
 Fig.~\ref{deltachi2} presents $\Delta \chi^2$ along some random samples of
 eigenvector directions and eigenvalues, $k=1,7,9,15$.
 The curves for middle values of $\lambda_{k=7,9}$ are very close to the
 ideal quadratic curve $\Delta \chi^2=t^2$ and for other
 eigenvalues $\lambda_{k=1,15}$ we see some departure of ideal
 quadratic curve which shows the quadratic approximation is almost adequate, though imperfect.
 The behavior of the some odd curves, which are also available in other QCD analysis \cite{Martin:2009iq}, usually
 correspond to the parameters controlling some unknown $x$-dependance parts of sea quarks and
 gluon densities.
\subsection{The spin sum rule}
In the framework of QCD the spin~$\frac{1}{2}$ of the proton can be
defined in terms of the first moment of the total quark and gluon
polarized densities and their orbital angular momentum
\begin{equation} \frac{1}{2}=\frac{1}{2}\Delta\Sigma^{p}+\Delta
g^{p}+L_{z}^{p}\;,\end{equation}
 where $L_{z}^{p}$ contains the total orbital angular momentum of all partons.
  The contribution of $\frac{1}{2}\Delta\Sigma+\Delta g$ in the scale
  of $Q^{2}=4$ GeV$^{2}$ is around $0.010$ in the present analysis. The
reported values from DSSV09 \cite{deFlorian:2009vb} and LSS10
\cite{lss} are $0.026$ and $-0.212$ respectively. For the positive
gluon scenario LSS obtained $0.419$ while our esult is $0.367$. In
Table \ref{sumrule} we compare the values of polarized PDFs first
moment in NLO approximation with other recent analyses. Since the
values of $\frac{1}{2}\Delta\Sigma$ are almost comparable, we
observe and conclude that the difference between the reported values
of $\frac{1}{2}\Delta\Sigma+\Delta g$ must be caused by different
gluon distributions. Indeed, proliferation in PPDFs data for sea
quarks from SIDIS experiments eventuates to more accurate results
for gluon distribution rather than analysis on DIS data merely, so
one cannot yet come to a definite conclusion about the contribution
of the orbital angular momentum to the total spin of the proton.
 \begin{table}[!ht]
\centering {\begin{tabular}{|cccc|}
\hline Fit & $\Delta \bar{s}$ & $\Delta G$ & $\Delta \Sigma$ \\
\hline
  DSSV09  & -0.056 & -0.096 & 0.245 \\
 LSS10   & -0.063 & ~0.316& 0.207 \\
  LSS10 ($\delta g$ pos)   & -0.055 & ~0.339& 0.254 \\
  MODEL  & -0.042 &-0.138& 0.256 \\
  MODEL ($\delta g$ pos)  & -0.046 &0.245& 0.244 \\
  \hline
\end{tabular}
} \caption{First moments of polarized PDFs at $Q^2=4$ GeV$^2$. The
corresponding DSSV09 and LSS10 values are also presented.
}\label{sumrule}
\end{table}
 The
estimation of the valence spin distribution can be written as an
accurate relation obtained from inclusive interactions in the
experiments. Indeed one obtains the following at LO
\cite{Alekseev:2007vi}
\begin{equation}
  \delta u_v + \delta d_v \sim \frac{36}{5} \frac{g_1^d}{(1 - 1.5
  \omega_D)} \ ,
  \label{eq:DIS_g1d}
\end{equation}
which is approximately true at NLO.

\begin{table*}[!ht]
%\hspace{-2cm}
\footnotesize{
\begin{tabular}{|l||c| c||c|c|c|c|c|c|}
\hline
 & $x$-range & $Q^2$ & \multicolumn{3}{c|}{$\Delta u_v + \Delta d_v$} & \multicolumn{3}{c|}{$\Delta\bar{u} + \Delta\bar{d}$} \\
\cline{4-9}
   &  & $\!\!\hspace{2mm} $(GeV$^2)\hspace{2mm}\!\!$       & Exp.Value  & DNS   &  MODEL  &    Exp.Value       & DNS &MODEL\\
\hline \hline
 SMC98 & $0.003-0.7$ & 10 & $0.26\pm0.21\pm0.11$ & 0.386 &0.454 & ~~$0.02\pm0.08\pm0.06$ & $-0.009$ & -0.043\\
\hline
 HERMES05 & $0.023-0.6$ & 2.5 & $0.43\pm0.07\pm0.06$ & 0.363 &0.445 & $-0.06\pm0.04\pm0.03$ & $-0.005$&-0.040\\
\hline
         & $0.006-0.7$ &     & $0.40\pm0.07\pm0.06$  & 0.385 &  0.459 &       --             & $-0.007$ &-0.043\\
 \raisebox{1.5ex}[-1.5ex]{COMPASS07}&       $0-1$ & \raisebox{1.5ex}[-1.5ex]{10} & $0.41\pm0.07\pm0.06$  &  --&--   & ~$0.0\pm0.04\pm0.03$ & --  &--\\
\hline
\end{tabular}}
\caption{Evaluations of the  first moments $\Delta u_v + \Delta d_v$
and $\Delta\bar{u} + \Delta\bar{d}$ from SMC \cite{SMC98}, HERMES
\cite{hermespdf} and COMPASS \cite{Alekseev:2007vi} data and also
from the DNS analysis \cite{deFlorian:2005mw} and present model
truncated to the
  range of each relevant experiment.
  The SMC results were obtained with the assumption of a SU(3) symmetric sea:
  $\Delta\bar{u}=\Delta\bar{d}=\Delta\bar{s}$.}
\label{FirstMom}
\end{table*}
 Theoretically, the first
moment of the polarized valence distributions, truncated to the
measured range of $x$,
\begin{equation}
\Gamma_v(x_{\rm min} < x < x_{\rm max}) = \int_{x_{\rm min}}^{x_{\rm
max}} \left[\delta u_v(x) + \delta d_v(x)\right] {\rm d}x \ ,
\label{eq:momment}
\end{equation}
is obtained and shown for current model and DNS
\cite{deFlorian:2005mw} in Table.~\ref{FirstMom}. We obtain for the
full measured range of $x$ in SMC \cite{SMC98}, HERMES
\cite{hermespdf} and COMPASS  \cite{Alekseev:2007vi}  experiments
\begin{eqnarray}
\Gamma_v(0.003 < x < 0.7) &=& 0.454 \ ,\\
\Gamma_v(0.023 < x < 0.6) &=& 0.445 \ ,\\
 \Gamma_v(0.006 < x < 0.7) &=&0.459 \ ,
\end{eqnarray}
at $Q^2 = 10,~2.5$ and $10$ GeV$^2$ respectively. Our value of
$\Gamma_v$ confirms the experimental results and the values come
from DNS analysis.

 An  estimation of the sea quark first moment
contribution to the nucleon spin can be generated by combining the
values of $\Gamma_v$, $\Gamma_1^N$ and $a_8$ \cite{Alekseev:2007vi}
\begin{equation}
\Delta {\overline u} + \Delta {\overline d} = 3 \Gamma_1^N -
\frac{1}{2} \Gamma_v + \frac{1}{12} a_8\ ,
\end{equation}
where $\Gamma_1^N$ is defined as the first moment of polarized
structure function $g_1$ for the average nucleon $N$ in an isoscalar
target: $g_1^N = (g_1^p + g_1^n)/2$
\begin{equation}
  \Gamma_1^N(Q^2\!=\! 10~{\rm GeV}^2) =
  \int_0^1 g_1^N(x,Q^2)\ ,
\label{gam}
\end{equation}
and ($a_8 = 3 F - D$) evaluated from semi-leptonic hyperon decays.
The result is reported to be zero for COMPASS experiment as shown in
Table~\ref{FirstMom}. The zero value of $\Delta {\overline u} +
\Delta {\overline d}$ suggests that $\Delta {\overline u}$ and
$\Delta {\overline d}$, if they are not zero, must be in opposite
sign. Previous estimation by SMC and HERMES are comparable with this
supposition and are also given in Table~\ref{FirstMom}. The DNS
parametrization, like the present model, finds a positive value for
$\Delta {\overline u}$ and a negative value for $\Delta {\overline
d}$, almost equal in absolute value. Opposite signs of $\Delta
{\overline u}$ and $\Delta {\overline d}$ are anticipated in several
models, e.g.\ in Ref.~\cite{deFlorian:2009vb,lss,Gluck:2000dy}, see
also \cite{peng} and references inside.

\section{Summary\label{summary}}
We have presented a NLO QCD analysis of the polarized DIS and SIDIS
data on nucleon. During the analysis we consider SU(2) and SU(3)
symmetry breaking scenario i.e. $\delta \bar{u}\neq\delta
\bar{d}\neq\delta\bar s$, since the available experimental data are
not enough to distinguish $\delta s$ from $\delta\bar{s}$, we take
them equal $\delta s=\delta \bar{s}$. The role of the semi-inclusive
data in determining the polarized sea quarks is discussed and we
have found also that the polarized gluon density is still ambiguous,
this ambiguity is the main reason that the quark-gluon contribution
into the total spin of the proton is still not well determined. We
also calculate the error of PPDFs by standard Hessian method and
investigate the quadratic behavior of $\Delta\chi^2$. Having
extracted the polarized PDFs, we compute the first moments of them
and discuss about sum rules.In general, we find good agreement with
the experimental data, and our results are in accord with other
determinations specially DSSV09 and LSS10 which are the most precise
ones. In conclusion, this demonstrates progress of the field toward
a detailed description of the spin structure of the nucleon \cite
{Soffer:2011wb,Shevchenko:2011nq,Leader:2010gz,deFlorian:2010aa,Sissakian:2009ag,Boer:2011fh,deFlorian:2012wk}.
The results of our new analysis applying pp collision data and
studying the impact of fragmentation functions on PPDFs will be
presented in a separate publication very soon.

\section{Acknowledgments}
\vspace{-0.2cm}We thank F. Olness for useful discussions and reading
the manuscript. F.A appreciates M. Stratmann from DSSV for kind
orientation and E. Kabuss from COMPASS collaboration for
communications. A.N.K thanks SITP (Stanford Institute for
Theoretical Physics) for partial support and the Physics Department
of SMU (Southern Methodist University) for their hospitality during
the completion of this work. He is also grateful to CERN TH-PH
division for the hospitality where a portion of this work was
performed. This project was financially supported by Semnan
University, Semnan University Science and Technology Park and the
School of Particles and Accelerators, Institute for Research in
Fundamental Sciences (IPM).
\appendix
\section{Fortran code}
A \texttt{FORTRAN} package containing our polarized parton
distributions $x\delta u_{v}(x,Q^{2})$, $x\delta d_{v}(x,Q^{2})$,
$x\delta g(x,Q^{2})$, $x\delta \bar{u}(x,Q^{2})$, $x\delta
\bar{d}(x,Q^{2})$ and $x\delta s(x,Q^{2})$ at NLO in the
$\overline{{\rm MS}}$--scheme can be found in
\texttt{http://particles.ipm.ir/links/QCD.htm} \cite{code} or
obtained via e-mail from the authors. These functions are
interpolated using cubic splines in $Q^{2}$ and a linear
interpolation in $\log\,(Q^{2})$. The package includes an example
program to illustrate the use of the routines.
\section{N-moment of all polarized SIDIS Wilson coefficients }
The transformation of Eq.~\ref{MellinC} gives the SIDIS Wilson
coefficients in $N-z$ space
\begin{eqnarray}\nonumber
 C_{qq}(N,z)&=&C_f\left\{-8\delta(1-z)+\tilde{P}_{qq}(z)\ln
\frac{Q^2}{M^2}+L_1(z)\right. \nonumber\\
&&+L_2(z)+(1-z) +\delta(1-z)\left[\right.\left(\right.\frac{1}{N}+
\frac{1}{1+N}\nonumber\\
&&-2\gamma-2\Psi(N)-\frac{2}{N}-\frac{2}{N+1}
+\frac{3}{2}\left.\right)\ln\frac{Q^2}{M^2} \nonumber\\
&&+\frac{1}{6}\left(\right.\frac{3}{N^2}+\frac{3}{(1+N)^2}
+6\gamma(\gamma+\frac{1}{N} +\frac{1}{1+N})\nonumber\\
&&+\pi^2
+3\Psi(N)\times(4\gamma+\Psi(N))+3\Psi^{2}(N+2)\nonumber\\
&&-6\frac{d\Psi(N)}{d N}\left.\right)
-\left(\right.-\frac{1}{N^2}+\frac{1}{(1+N)^2}\nonumber\\
&&-2\zeta(2,N+1)\left.\right)+\frac{1}{N+N^2}\left.\right]\nonumber\\
&&-2(\gamma+\Psi(N))\frac{1}{(1-z)_+}+(1+z)(\gamma+\Psi(N))\nonumber\\
&&-(\frac{1}{N}+\frac{1}{1+N})\frac{1}{(1-z)_+} \nonumber\\
&&\left.+\frac{2(1+N+NZ)}{N+N^2}-\frac{2(1-z)}{N+N^2}\right\}~,
\end{eqnarray}
\begin{eqnarray}\nonumber
C_{gq}(N,z)&=&C_f\left\{\frac{1+(1-z)^2}{z}\left[\ln \left(
\frac{Q^2}{M^2}z(1-z)\right) -\gamma
\right]\right.\nonumber\\&&-\Psi(N) +z+2\frac{1+2N-N z}{N+N^2}
-\frac{1+2N}{N z +N^2z}\nonumber\\
&&\left.-\frac{2z}{N+N^2}\right\}~,
\end{eqnarray}
\begin{eqnarray}
C_{qg}(N,z)&=&\frac{1}{2}\left\{\delta(1-z)\right.\nonumber\\
&&\times\left[-\frac{2-2N-(2+N+N^2)
(\gamma+\Psi(N))}{N(1+N)(2+N)}\right.\nonumber\\
&&+\left.\left.\frac{2}{2+3N+N^2}\right]\right.\nonumber\\
&&+\frac{2+N+N^2}{2N+3N^2+N^3}
\left[\frac{1}{(1-z)_+}\right.\left.\left.+\frac{1}{z}-2\right]\right\}~,\nonumber\\
\end{eqnarray}
\begin{eqnarray}
 \delta C_{qq}(N,z)&=&C^1_{qq}-2C_f\frac{1}{N+N^2}(1-z)~,\\
 \delta C_{qq}(N,z)&=&C^1_{qq}-2C_f\frac{1}{N+N^2}z~,\\
\delta
C_{qg}(N,z)&=&\frac{1}{2}\left\{\delta(1-z)\left[-\frac{2+(N-1)
(\gamma+\Psi(N))}{N(1+N)}\right.\right.\nonumber\\
&&\left.+\frac{2}{N+N^2}\right]\nonumber\\
&&+\frac{N-1}{N(1+N)}\left[\frac{1}{(1-z)_+}\right.\left.\left.+\frac{1}{z}-2\right]\right\}~,
\end{eqnarray}
where $C_f=4/3$, $\gamma\simeq 0.577216$ and
\begin{eqnarray}
\tilde{P}_{qq}(z)&=&\frac{1+z^2}{(1-z)_+}+\frac{3}{2}\delta(1-z)~,\\
L_1(z)&=&(1+z^2)\left(\frac{\ln(1-z)}{1-z}\right)~,\\
L_2(z)&=&\frac{1+z^2}{1-z}\ln z~,\\
\int_0^1 dz~f(z)(g(z))_{+}&\equiv&\int_0^1dz[f(z)-f(1)]g(z).
\end{eqnarray}


\end{document}